\documentclass[12pt]{article}
\usepackage{graphicx,amsmath,cite}

\newcommand{\KaTie}{{\sc Ka\nolinebreak\hspace{-0.3ex}Tie}}

\newcommand\pubnumber{IFJPAN-IV-2017-22}
\newcommand\pubdate{\today}

\def\napoli{Institute of Nuclear Physics Polish Academy of Sciences,\\ PL-31342 Krak\'ow, Poland}
\def\support{\footnote{This work was supported by grant No.~2015/17/B/ST2/01838 of the National Science Center, Poland.}}

\def\Title#1{\begin{center} {\Large #1 } \end{center}}
\def\Author#1{\begin{center}{ \sc #1} \end{center}}
\def\Address#1{\begin{center}{ \it #1} \end{center}}

\newcommand\pubblock{\rightline{\begin{tabular}{l}
         \pubnumber\\
         \pubdate  \end{tabular}}}
\newenvironment{Abstract}{\begin{quotation}  }{\end{quotation}}
\newenvironment{Presented}{\begin{quotation} \begin{center} 
             PRESENTED AT\end{center}\bigskip 
      \begin{center}\begin{large}}{\end{large}\end{center} \end{quotation}}

\def\beq{\begin{equation}}
\def\eeq#1{\label{#1}\end{equation}}
\def\eeqn{\end{equation}}

\def\beqa{\begin{eqnarray}}
\def\eeqa#1{\label{#1}\end{eqnarray}}
\def\eeqan{\end{eqnarray}}

\let\bar=\overbar

\def\eg{{\it e.g.}}

\def\Dslash{\not{\hbox{\kern-4pt $D$}}}
\def\dslash{\not{\hbox{\kern-2pt $\del$}}}

\def\msb{{\bar{\ssstyle M \kern -1pt S}}}

\begin{document}
\begin{titlepage}
\pubblock

\vfill
\Title{One-loop amplitudes with an off-shell gluon}
\vfill
\Author{A.~van~Hameren\support}
\Address{\napoli}
\vfill
\begin{Abstract}
I report on recent advancements regarding calculations that require initial-state partons with non-vanishing transverse momentum, causing it to be space-like and thus off-shell. 
\end{Abstract}
\vfill
\begin{Presented}
Presented at EDS Blois 2017, Prague, \\ Czech Republic, June 26-30, 2017
\end{Presented}
\vfill
\end{titlepage}
\def\thefootnote{\fnsymbol{footnote}}
\setcounter{footnote}{0}

\section{Introduction}
Factorization prescriptions that involve an explicit dependence on the transverse momentum ($k_T$) of one or both initial-state partons generally require partonic hard matrix elements with space-like initial-state partons.
Examples are $k_T$-factorization~\cite{Gribov:1984tu}, high-energy factorization~\cite{Catani:1990eg,Collins:1991ty}, and a recent example is improved transverse momentum dependent factorization~\cite{Kotko:2015ura}.
The definition and calculation of the partonic matrix elements needs some special care in order to insure gauge invariance, but several approaches exist that lead to equivalent results at tree-level~\cite{Lipatov:1995pn,Lipatov:2000se,Antonov:2004hh,vanHameren:2012if,vanHameren:2013csa,Kotko:2014aba,Karpishkov:2017kph}, and even allow for so-called ``on-shell recursion'' for amplitudes, leading to compact expressions remarkably similar to those for fully on-shell amplitudes~\cite{vanHameren:2014iua,vanHameren:2015bba}.
Tree-level calculations have been completely automated in the parton-level Monte Carlo program \KaTie~\cite{vanHameren:2016kkz}, which can be used in combination with TMDlib~\cite{Hautmann:2014kza} to produce fully exclusive parton-level event files for arbitrary processes within the Standard Model.
Recently, off-shell amplitudes also were applied in the high-energy resummation of jet observables~\cite{Zoia:2017vfn}, in the context of Wilson line and local operators in $\mathcal{N}=4$ Super Yang-Mills theory~\cite{Bork:2017qyh}, and appeared in the context of light-front quantization and the MHV action~\cite{Kotko:2017nkx}.

The obvious next step is to go beyond leading-order perturbation theory, and to perform calculations at next-to-leading order (NLO).
This is necessary both to reach higher precision and to assess the reliability of the factorization procedure.
It must allow for a systematic treatment of the divergences that accompany higher-order calculations within quantum chromodynamics in order to be physically relevant.
Some results exist within the so-called parton reggeization approach~\cite{Hentschinski:2011tz,Chachamis:2012cc,Chachamis:2013hma,Nefedov:2016clr,Nefedov:2017qzc}.
Other recent NLO calculations with explicit $k_T$ dependence are~\cite{Iancu:2016vyg,Boussarie:2016ogo,Ducloue:2017mpb,Beuf:2017bpd}.
An important ingredient of an NLO calculation is the one-loop amplitude, and complications arise with respect to space-like initial-state partons regarding {\em linear denominators\/}.
They lead to divergencies that cannot be straightforwardly regularized with dimensional regularization.
Recently, a regularization was proposed~\cite{vanHameren:2017hxx} that comes naturally with the approach to define off-shell amplitudes from~\cite{vanHameren:2012if}.
It manifestly respects Lorentz covariance, gauge invariance, and allows for practical calculations for arbitrary partonic processes.

\section{KaTie}
\KaTie\ is a parton-level event generator in the spirit of HELAC~\cite{Cafarella:2007pc} and {\sc Alpgen}~\cite{Mangano:2002ea}, with the extra feature that it allows for an explicit non-vanishing transverse momentum for the initial-state partons.
It can be downloaded from
\begin{center}
{\tt https://bitbucket.org/hameren/katie/downloads/}
\end{center}
It can generate event files in the LHEF format~\cite{Alwall:2006yp}, or in a custom file format.
For the latter, \KaTie\ also provides the necessary histogram routines to plot arbitrary distributions.
\KaTie\ requires LHAPDF~\cite{Buckley:2014ana} for collinear PDFs and for the running coupling.
Transverse momentum dependent PDFs (TMDPDFs) can be provided by TMDlib~\cite{Hautmann:2014kza}, or as ASCII files containing hyper-rectangular grids which \KaTie\ itself interpolates.
A hadron-level process, for example $pp\to b\bar{b}\,\mu^+\mu^{-}$, is completely defined in a single input file, which must contain
\begin{itemize}
\item the list of desired partonic subprocesses, \eg\\\hspace*{8ex}
$g^*g^*\to b\bar{b}\,\mu^+\mu^{-}$\quad,\quad
 $u^*\bar{u^*}\to b\bar{b}\,\mu^+\mu^{-}$\quad,\quad
 $d^*\bar{d^*}\to b\bar{b}\,\mu^+\mu^{-}$ ;
\item the identifier for the collinear PDF set in LHAPDF;
\item the identifier for the TMDPDF set in TMDlib, or alternatively a list of grid files and the directory where they can be found;
\item the value of the center-of-mass energy and a list of phase space cuts;
\item the desired values of couplings and masses and widths of the particles involved;
\item the desired interaction, \eg\ both QCD and electroweak interactions for this example.
\end{itemize}
Running the program happens in two stages: in the first (short) stage, the phase space pre-sampler is optimized for each partonic sub-process separately.
This stage is not parallelizable, but should in general be rather quick.
The second stage is the actual event generation.
This can easily be parallelized by running several instances of an executable with different random number seeds.

\newcommand{\imag}{\mathrm{i}}
\newcommand{\lid}[2]{#1\!\cdot\!#2}
\newcommand{\slashl}{\ell\hspace{-6pt}/}
\newcommand{\slashp}{p\hspace{-5.5pt}/}
\newcommand{\slashq}{q\hspace{-6.5pt}/}
\newcommand{\slashk}{k\hspace{-6.5pt}/}
\newcommand{\slashK}{K\hspace{-8.5pt}/}
\newcommand{\slashe}{e\hspace{-6.0pt}/}
\newcommand{\slashr}{r\hspace{-6.0pt}/}
\newcommand{\Amp}{\mathcal{A}}
\section{One-loop amplitudes}
Tree-level amplitudes with off-shell gluons were defined in \cite{vanHameren:2012if} by embedding the process into an on-shell scattering process with an auxiliary ``flavor $A$'' quark-antiquark pair instead of the off-shell gluon.
The momenta $p_A^\mu,p_{A'}^\mu$ of the auxiliary quark-antiquark pair add up to the desired momentum $k^\mu=xp^\mu+k_T^\mu$ for the off-shell gluon, where $p^\mu$ is the momentum of one of the scattering hadrons.
This can be achieved for example as follows:
%
%%%%%%%%%%%%%%%%%%%%%%%%%%%%%%%%%%%%%%%%
\begin{equation}
p_A^\mu = \Lambda p^\mu + \alpha q^\mu + \beta k_{T}^\mu
\quad,\quad
p_{A'}^\mu =xp^\mu+k_T^\mu-p_A^\mu %-(\Lambda-x) p^\mu - \alpha q^\mu + (1-\beta) k_{T}^\mu
~,
\label{auxmom1}
\end{equation}
%%%%%%%%%%%%%%%%%%%%%%%%%%%%%%%%%%%%%%%%
%
where $q^\mu$ is an arbitrary light-like momentum with $p\!\cdot\!q>0$, where $q\cdot k_T=p\cdot k_T=0$, and where $\alpha =-\beta^2k_{T}^2/\big(\Lambda(p+q)^2\big)$ with $\beta =1/\big(1+\sqrt{1-x/\Lambda}\big)$.
With this choice, the momenta $p_A,p_{A'}$ satisfy the desired relations
%
%%%%%%%%%%%%%%%%%%%%%%%%%%%%%%%%%%%%%%%%
\begin{equation}
p_A^2 = p_{A'}^2 = 0
\quad,\quad
p_A^\mu+p_{A'}^\mu = xp^\mu+k_T^\mu
\end{equation}
%%%%%%%%%%%%%%%%%%%%%%%%%%%%%%%%%%%%%%%%
%
for any value of the parameter $\Lambda$.
The amplitude with the off-shell gluon is obtained by taking $\Lambda\to\infty$:
%
%%%%%%%%%%%%%%%%%%%%%%%%%%%%%%%%%%%%%%%%
\begin{equation}
\frac{|k_T|}{\Lambda}\,
\Amp\Big(\emptyset\to \bar{\mathrm{q}}_A\big(p_A(\Lambda)\big)\,\mathrm{q}_A\big(p_{A'}(\Lambda)\big)+X\Big)
\;\overset{\Lambda\to\infty}{\longrightarrow}\;
\Amp\big(\emptyset\to g^*(xp+k_T)+X\big)
~.
\end{equation}
%%%%%%%%%%%%%%%%%%%%%%%%%%%%%%%%%%%%%%%%
%
Here $X$ stands for other particles in the partonic process, like $gb\bar{b}\mu^+\mu^-$ in the example of the previous section with only one off-shell gluon.
This limit is rather straightforward for tree-level amplitudes, because they are just rational functions of the external momenta, and the limit boils down to the limit
%
%%%%%%%%%%%%%%%%%%%%%%%%%%%%%%%%%%%%%%%%
\begin{equation}
\imag\,\frac{\slashp_A+\slashK}{(p_A+K)^2}
\overset{\Lambda\to\infty}{\longrightarrow} \frac{\imag\,\slashp}{2p\!\cdot\!K}
\label{regulator}
\end{equation}
%%%%%%%%%%%%%%%%%%%%%%%%%%%%%%%%%%%%%%%%
%
for the auxiliary quark propagators.
Taking this limit on the integrand of a one-loop amplitude with an auxiliary quark-antiquark pair would lead to linear denominators of the type $2p\!\cdot\!K$ where $K^\mu$ contains the loop integration momentum $\ell^\mu$.
Such loop denominators lead to divergencies that cannot straightforwardly be regulated with dimensional regularization.
In~\cite{vanHameren:2017hxx} it was proposed to use $\Lambda$ as the regulator, and perform (\ref{regulator}) after the loop integration.
Before the limit, all loop denominators are quadratic, and the loop integrals are well-defined within dimensional regularization.
The divergencies eventually show up as powers of $\log\Lambda$.

It turns out that some complications arise due to the non-commutativity of counting powers of $\Lambda$ and performing the loop integral.
Consequently, care has to be taken in identifying all non-vanishing contributions in the limit $\Lambda\to\infty$.
Furthermore, it is not {\it a priori\/} clear that there are no contributions that diverge linearly or worse with $\Lambda$.
In \cite{vanHameren:2017hxx} it is shown, however, that the divergencies show up as at most $\log^2\Lambda$, and that the powerful {\em integrand methods\/} for one-loop integrals~\cite{Ossola:2006us,Ellis:2007br} can be applied to large extend to calculate the amplitudes.

%%%%%%%%%%%%%%%%%%%%%%%%%%%%%%%%%%%%%%%%%%%%%%%%%%%%%%%%%%%%%%%%%%%%%%%%%
%%
%%   use this format to include an .eps figure into your paper
%%
% \begin{figure}[htb]
% \centering
% \includegraphics[height=1.5in]{magnet}
% \caption{Plan of the magnet used in the mesmeric studies.}
% \label{fig:youfigure}
% \end{figure}
%%%%%%%%%%%%%%%%%%%%%%%%%%%%%%%%%%%%%%%%%%%%%%%%%%%%%%%%%%%%%%%%%%%%%%%%%%%

%%%%%%%%%%%%%%%%%%%%%%%%%%%%%%%%%%%%%%%%%%%%%%%%%%%%%%%%%%%%%%%%%%%%%%%%%
%%
%%   use this format to include a LaTeX table  into your paper
%%
% \begin{table}[t]
% \begin{center}
% \begin{tabular}{ccc}  
%  ...
% \end{tabular}
% \caption{Blood cyanide levels for the two patients.}
% \label{tab:blood}
% \end{center}
% \end{table}
%%%%%%%%%%%%%%%%%%%%%%%%%%%%%%%%%%%%%%%%%%%%%%%%%%%%%%%%%%%%%%%%%%%%%%%%%%%

\end{document}